\documentclass{elsart}

 \usepackage{graphicx}
\usepackage{amssymb}

\begin{document}

\begin{frontmatter}



\title{Effects of nuclear molecular configurations on the astrophysical 
S-factor for $^{16}$O + $^{16}$O}


\author[label1]{A. Diaz-Torres}
\ead{alexis.diaz-torres@anu.edu.au}
\author[label1]{, L.R. Gasques}
\author[label2]{and M. Wiescher}
\address[label1]{Department of Nuclear Physics, Research School of Physical
Sciences and Engineering, Australian National University, Canberra,
ACT 0200, Australia}
\address[label2]{Department of Physics and Joint Institute for Nuclear Astrophysics (JINA),
University of Notre Dame, Notre Dame, IN 46556, USA}

\begin{abstract}
% Text of abstract
The impact of nuclear molecular configurations on the astrophysical S-factor for
$^{16}$O + $^{16}$O is investigated within the realistic
two-center shell model based on Woods-Saxon potentials. These
molecular effects refer to the formation of a neck between the
interacting nuclei and the radial dependent collective mass
parameter. It is demonstrated that the former is crucial to
explain the current experimental data with high accuracy and
without any free parameter, whilst in addition the latter predicts a 
pronounced maximum in the S-factor. In contrast to very recent results by Jiang et al., 
the S-factor does not decline towards extremely low values as energy decreases.
\end{abstract}

\end{frontmatter}

% main text
%\section{}
%\label{}

Low-energy fusion reactions involving $^{12}$C and $^{16}$O are of
great astrophysical importance for our understanding of the
timescale and the nucleosynthesis during late stellar evolution.
Fusion processes like $^{12}$C+$^{12}$C, $^{12}$C+$^{16}$O, and
even $^{16}$O+$^{16}$O characterize the carbon burning phase of
massive stars (M$\ge$8M$_{\odot}$) with $^{16}$O+$^{16}$O being
the key reaction for the later oxygen burning phase of these stars
\cite{Rolfs}. The timescale for these burning phases depends on
the mass of the star as well as on the reaction rates of the
associated fusion processes. The nucleosynthesis during carbon and
oxygen burning depends not only on the reaction rate but also on
the branching between proton, neutron, or alpha decay channels of the
fused compound nucleus \cite{WAG2006}. In order to obtain
astrophysical reaction rates for different stellar burning
scenarios it is crucial to know the fusion cross sections at very
low energies. Usually these quantities are extrapolated from
theoretical calculations that explain the relatively 
high-energy fusion data because direct experiments at very low energies 
are extremely difficult to carry out. The S-factor \cite{Rolfs} is an
alternative representation of the fusion cross section ($S =
\sigma_{fus} E e^{2\pi \eta}$, where $\eta$ is the Sommerfeld
parameter) that facilitates this extrapolation, in particular, for
reactions with very light projectiles such as hydrogen or helium
because it changes slowly with energy. For heavier nuclei, the
S-factor may depend strongly on energy due to the higher Coulomb
barriers and angular momenta involved in the fusion process.

The aim of this letter is to show the impact of molecular effects
on the astrophysical S-factor for $^{16}$O + $^{16}$O. The
motivation is due to the current discussion in the literature
about the behavior of the S-factor excitation function at very low
incident energies. Does this function take very small values with
decreasing energy? Is there a maximum in the S-factor at energies
around 7-8 MeV, as a recent empirical analysis by Jiang et al.
\cite{Jiang} suggests? In this work we give a comprehensive answer
to these questions by studying the above reaction within the
realistic two-center shell model (TCSM) based on spherical
Woods-Saxon potentials, as recently reported in Ref.
\cite{SPH_TCWS}. The  $^{16}$O + $^{16}$O reaction is a good test
case due to its theoretical simplicity (the nuclei are spherical
and coupled channels effects are expected to be irrelevant).
Furthermore, abundant experimental sub-barrier fusion data
\cite{Exp1,Exp2,Exp3,Exp4,Exp5} exist for comparison to the model
calculations.

The molecular picture \cite{GreinerParkScheid} is justified at
sub-barrier energies since the radial motion of the nuclei is
expected to be adiabatically slow compared to the rearrangement of
the mean field of nucleons. We will show that molecular effects
related to the formation of a neck between the nuclei and the
radial dependent mass parameter are crucial to understand the
S-factor excitation function at low energies. Most current
theoretical studies use a sudden potential (such as the
double-folding potential) and a constant reduced mass (see e.g.
Ref. \cite{Gasques}). However, previous works
\cite{Flocard,Heenen,Urbano,Reinhard} based on either the double
oscillator potential or the adiabatic time dependent Hartree-Fock
theory have indicated that molecular aspects of the reaction are
very important. It is worth mentioning that some light heavy-ion
reactions of astrophysical interest have been studied by the
Frankfurt group within the concept of nuclear molecules
\cite{GreinerParkScheid} using a restricted TCSM constructed with
two harmonic-oscillator potentials.

%\begin{figure}
%\begin{center}
%\includegraphics[width=12.0cm]{DTCWS.eps}
%\end{center}
%\caption{Schematic picture of the coordinates used to define the two-center potential
%(\ref{eq1}) in the collision between two deformed nuclei. See text for further details.}
%\end{figure}

In the present work, the adiabatic collective potential energy
surface $V(R)$ is obtained with Strutinsky's
macroscopic-microscopic method, whilst the radial dependent
collective mass parameter $M(R)$ is calculated with the cranking
mass formula \cite{Inglis}. The rotational moment of inertia of
the dinuclear system is defined as the product of the cranking
mass and the square of the internuclear distance. The macroscopic
part of the potential results from the finite-ranged liquid drop
model with the parameters given in Ref. \cite{Moeller1981} and the
sequence of nuclear shapes generated with the TCSM
\cite{SPH_TCWS}. The microscopic shell corrections to the
potential are calculated with the novel method suggested in Ref.
\cite{LanczosDiaz}. The TCSM is used to calculate the neutron and
proton energy levels $E_i$ as a function of the separation $R$
between the nuclei along with the radial coupling \cite{SPH_TCWS}
between these levels that appears in the numerator of the cranking
mass expression,

\begin{equation}
M(R) = 2 \hbar^2 \sum_{i=1}^{A} \sum_{j > A}
\frac{|\langle j| \partial / \partial R |i \rangle|^2}
{E_j - E_i}.
\label{eq1}
\end{equation}

The parameters of the asymptotic WS potentials including the
spin-orbit term reproduce the experimental single-particle energy
levels around the Fermi surface of $^{16}$O \cite{SPH_TCWS},
whilst for $^{32}$S the parameters of the global WS potential by
Soloviev \cite{Soloviev} are used, its depth being adjusted to
reproduce the experimental neutron and proton separation energies
\cite{Audi}. To describe the amalgamation of two nuclei, the
potential parameters have to be interpolated between their values
for the separated nuclei and the compound nucleus. The parameters
can be correlated by conserving the volume enclosed by certain
equipotential surface of the two-center potential for all
separations $R$ between the nuclei (see Ref. \cite{SPH_TCWS} for
further details).

\begin{figure}
\begin{center}
\includegraphics[width=12.0cm,angle=0]{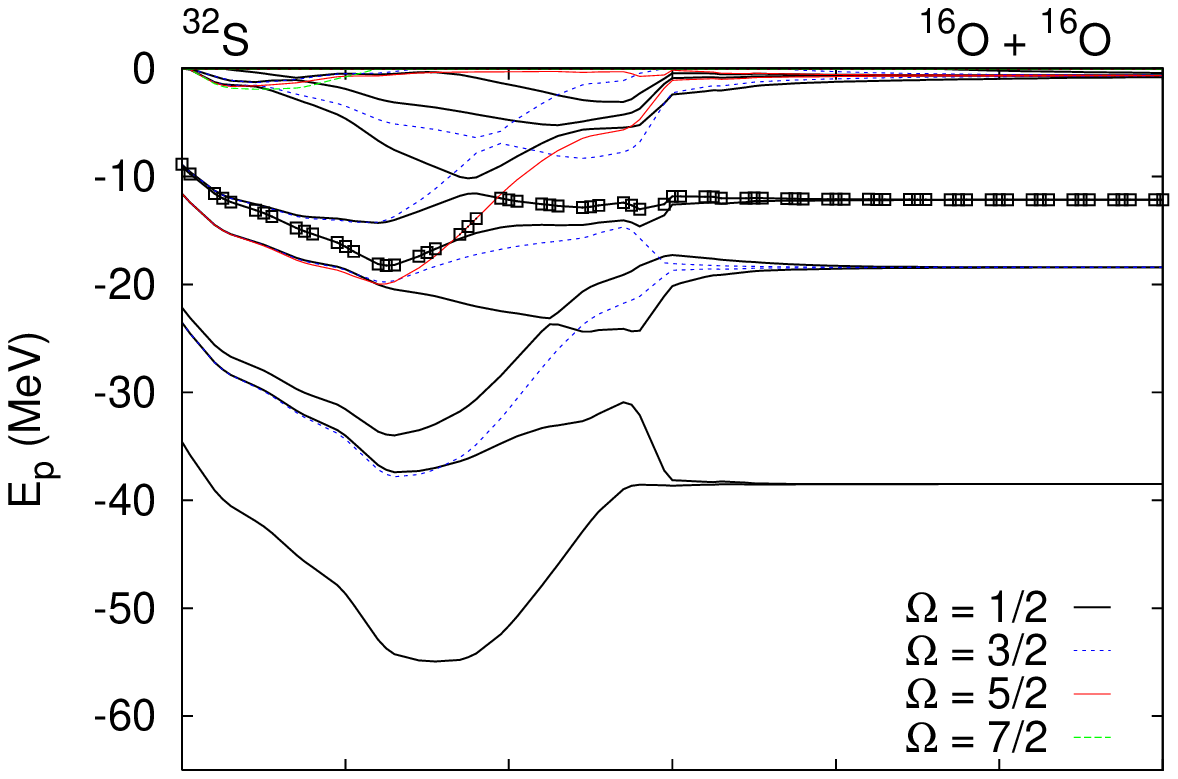} \\
\includegraphics[width=12.0cm,angle=0]{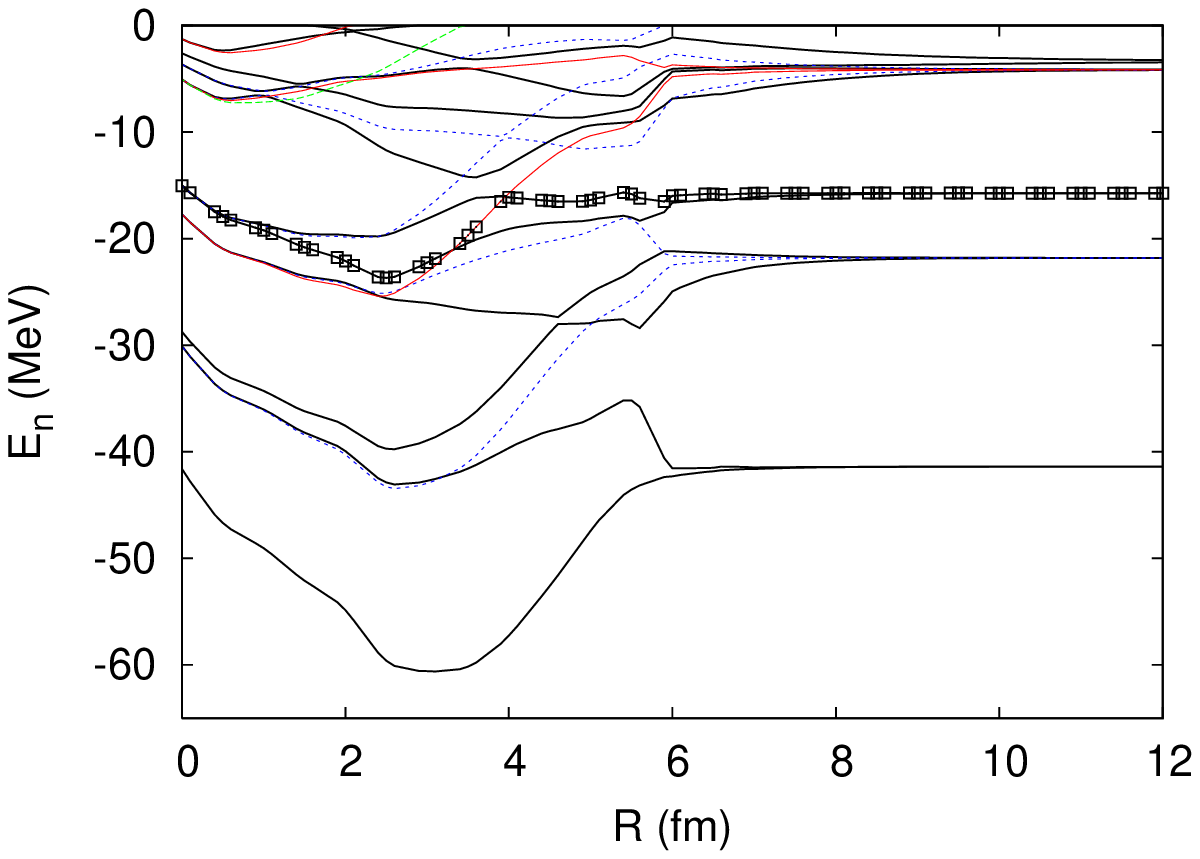}
\caption{(Color online) Molecular adiabatic single-particle levels
as a function of the separation between the nuclei: protons (top)
and neutrons (bottom). Different types of curve refer to states
with different projection of the nucleon total angular momentun
along the internuclear axis. The open squares denote the Fermi
surface. See text for further details.} \label{SPECTRA}
\end{center}
\end{figure}

\begin{figure}
\begin{center}
\includegraphics[width=12.0cm,angle=0]{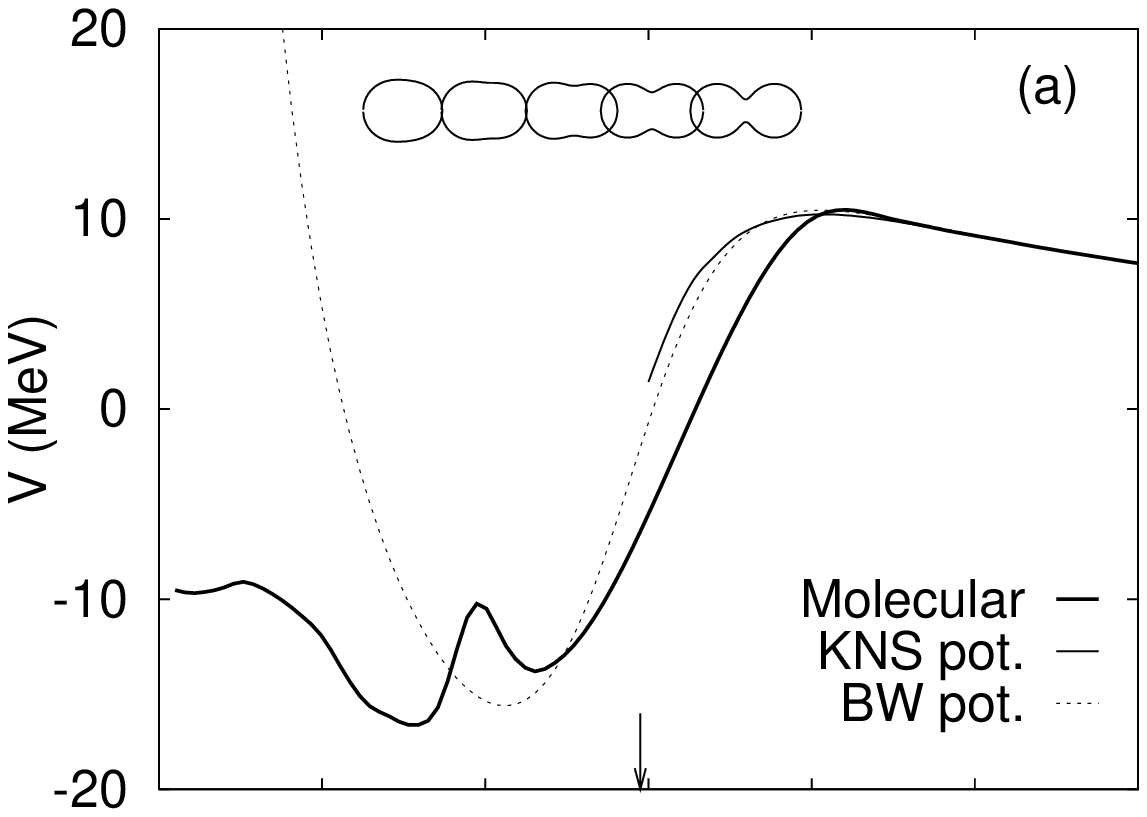} \\
\includegraphics[width=12.0cm,angle=0]{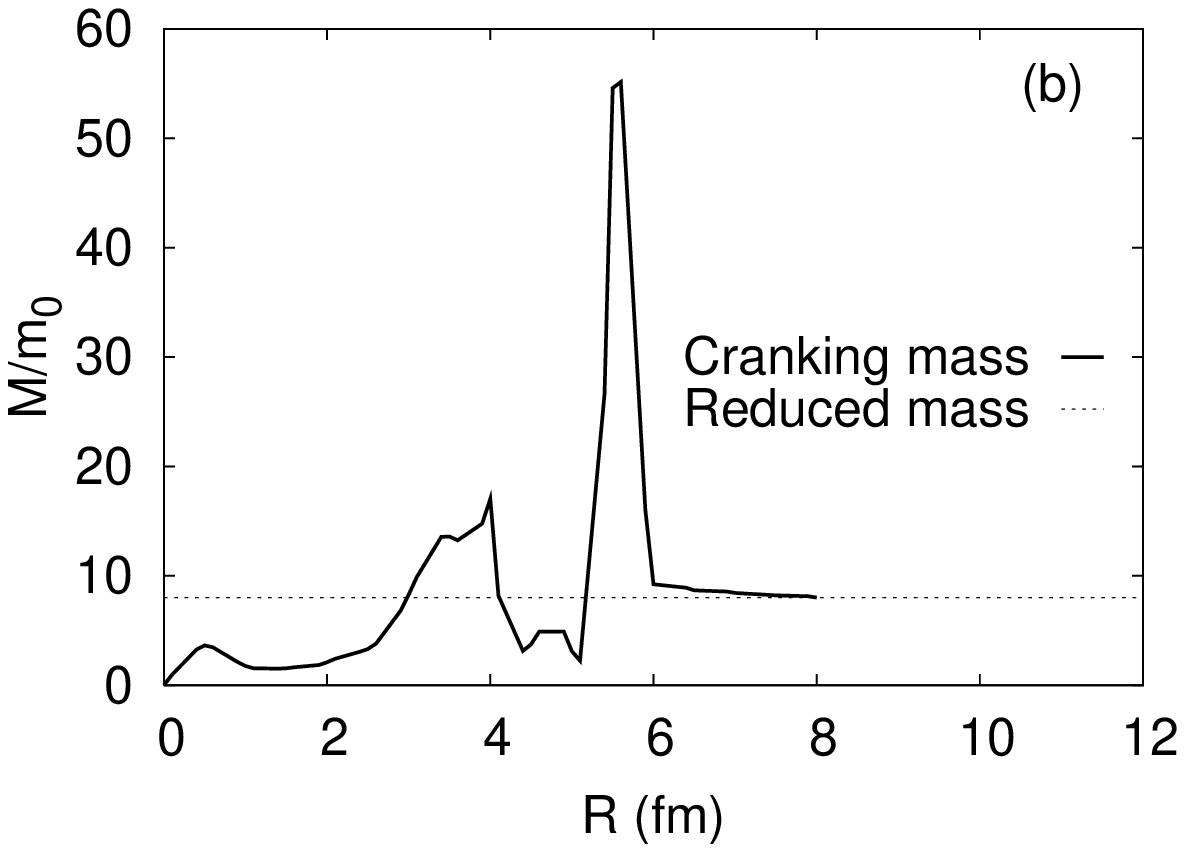}
\caption{(a) The s-wave collective potential energy as a function
of the separation between the nuclei for $^{16}$O + $^{16}$O. The
arrow indicates the geometrical contact separation. (b) The radial
dependent collective mass parameter (in units of nucleon mass
$m_0$). See text for further details.} \label{POTINERT}
\end{center}
\end{figure}

Fig. \ref{SPECTRA} shows the proton (top) and neutron (bottom)
molecular adiabatic levels as a function of the separation between
the nuclei. It can be observed that the shell structure of the
asymptotic nuclei essentially remains intact (very small
polarization of the energy levels) up to the geometrical contact
separation ($R_c= 5.85$ fm) that is well inside the s-wave capture
barrier (see Fig. \ref{POTINERT}a - arrow). For radii smaller than
$R_c$ (compact shapes) a significant rearrangement of the shell
structure of the fusing system occurs. This is reflected in the
collective potential energy surface and mass parameter, presented
in Fig. \ref{POTINERT}, by means of the shell correction energy
and the virtual excitation of the nucleons (induced by the radial
coupling between the single-particle states) to levels above the
Fermi surface (open squares), respectively.

Fig. \ref{POTINERT}a shows the s-wave molecular adiabatic
potential (thick solid curve) as a function of the internuclear
distance, which is normalized with the experimental $Q$-value of
the reaction ($Q=16.54$ MeV). The sequence of nuclear shapes
related to this potential \cite{SPH_TCWS} is also presented. For
comparison we show the Krappe-Nix-Sierk (KNS) potential \cite{KNS}
(thin solid curve) and the empirical Broglia-Winther (BW)
potential \cite{BW} (dotted curve). Effects of neck between the
interacting nuclei, before they reach the geometrical contact
separation, are not incorporated into the KNS potential. The
concept of nuclear shapes is not embedded in the BW potential
which tends to be similar to the KNS potential. Comparing the 
KNS potential to the molecular adiabatic potential we note that the neck
formation substantially decreases the potential energy after
passing the barrier radius ($R_b = 8.4$ fm). It will be shown that
the inclusion of neck effects is crucial to successfully explain the 
available S-factor data \cite{Exp1,Exp2,Exp3,Exp4,Exp5} for the 
studied reaction.

Fig. \ref{POTINERT}b shows the radial dependent cranking mass
(thick solid curve), whilst the asymptotic reduced mass is
indicated by the dotted line. Just passing the barrier radius,
when the neck between the nuclei starts to develop, the cranking
mass slightly increases compared to the reduced mass and
pronounced peaks appear inside the geometrical contact separation.
For the studied reaction, these peaks are mainly caused by the
strong change of the single-particle wave functions during the
rearrangement of the shell structure of the asymptotic nuclei into
the shell structure of the compound system [large values of the
radial coupling in the numerator of the cranking mass formula
(\ref{eq1})]. In general, the peaks could also be due to avoided
crossings \cite{SPH_TCWS} between the adiabatic molecular 
single-particle states (see Fig. \ref{SPECTRA}), which can make 
the denominator of the cranking mass expression (\ref{eq1}) very small.

\begin{figure}
\begin{center}
\includegraphics[width=12.0cm]{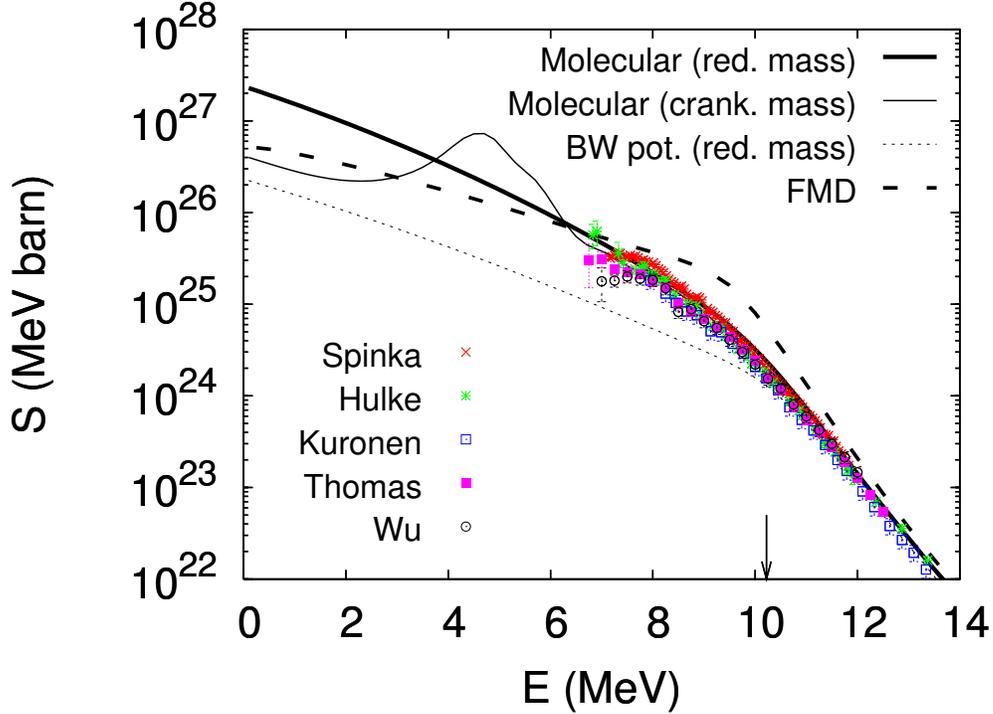}
\end{center}
\caption{(Color online) The S-factor as a function of the
center-of-mass energy for $^{16}$O + $^{16}$O. The curves are
theoretical calculations, whilst the symbols refer to experimental
data. The arrow indicates the Coulomb barrier of the molecular
potential of Fig \ref{POTINERT}a. See text for further details.}
\label{SFACT}
\end{figure}

Having the adiabatic potential and the adiabatic mass parameter,
the radial Schr\"odinger equation is exactly solved with the
modified Numerov method \cite{Melkanoff} and the incoming wave
boundary condition imposed inside the capture barrier. The fusion 
cross section $\sigma_{fus}$ is calculated taken into account the 
identity of the interacting nuclei and the parity of the wave function 
for the relative motion (only even partial waves $L$ are included here),
i.e., $\sigma_{fus} = \pi \hbar^2/(2 \mu E) \sum_{L}
(2L+1)(1+\delta_{1,2})P_L$, where $\mu$ is the asymptotic reduced
mass, $E$ is the incident energy in the total center-of-mass
reference frame and $P_L$ is the partial tunneling probability.

\begin{figure}
\begin{center}
\includegraphics[width=12.0cm]{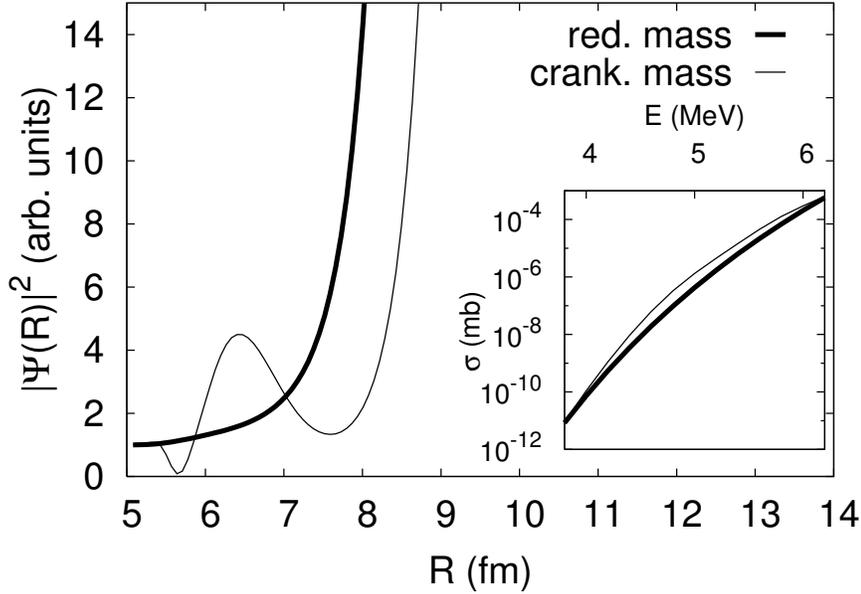}
\end{center}
\caption{Probability distribution of the collective radial wave function for the s-wave 
calculated with the reduced mass and the cranking mass at an incident energy of 4.5 MeV. 
The strong peak of the cranking mass in Fig. \ref{POTINERT}b makes it most probable for 
the molecular configuration (trapped inside the barrier) to be localized just before the 
geometrical contact separation (6-7 fm), which increases the fusion cross section 
(small figure inserted) and causes the local maximum of the S-factor in Fig. \ref{SFACT}. 
See text for further details.}
\label{WF}
\end{figure}

Fig. \ref{SFACT} shows the S-factor as a function of the incident
energy in the center-of-mass reference frame. For a better
presentation, the experimental data of each set
\cite{Exp1,Exp2,Exp3,Exp4,Exp5} are binned into $\Delta E = 0.5$
MeV energy intervals. In this figure the following features can be
observed:
\begin{description}
\item [\textnormal{(i)}] the molecular adiabatic potential of Fig.
\ref{POTINERT}a correctly (thick and thin solid curves) explains
the measured data, in contrast to either the results
obtained with the BW potential (dotted curve) or the very recent 
calculations by Neff et al. \cite{Neff} within the Fermionic Molecular 
Dynamics (FMD) approach (dashed curve). Since the width of the barrier 
decreases for the molecular adiabatic potential of Fig. \ref{POTINERT}a, 
it produces larger fusion cross sections than those arising from the 
shallower KNS and BW potentials. 
\item [\textnormal{(ii)}] the use
of the cranking mass parameter of Fig. \ref{POTINERT}b notably
affects the low energy S-factor, which is revealed by the
comparison between the thick and thin solid curves. It starts
reducing the S-factor around 7-8 MeV energy region and produces a
local maximum around 4.5 MeV. At the lowest incident energies (below
4 MeV) the S-factor is suppressed by a factor of five compared to
that arising from a constant reduced mass. The peak in the S-factor 
is due to an increase of the fusion cross section (see Fig. \ref{WF}), which is 
caused by the resonant behavior of the collective radial wave function increasing 
the fusion transmission coefficient. 
\end{description}

Since we have solved a single-channel Schr\"odinger equation (elastic channel) with the 
incoming wave boundary condition, which is equivalent to a very strong absorption inside 
the barrier, we do not see any molecular resonances for the lowest partial waves leading 
to fusion. However, few of them may exist for the grazing waves which should be reflected 
in direct reaction channels \cite{GaulBickel}. At incident 
energies higher than those studied in the paper (i.e., E $\gtrsim$ 20 MeV), 
we expect that strong inelastic channels will be open, and the molecular resonances could 
impact not only on the elastic, inelastic and transfer excitation functions but also on 
the fusion excitation function. Here a molecular Coupled Reaction Channels (CRC) description 
\cite{GreinerParkScheid} will be required. 

In summary, the adiabatic molecular potential appears to be crucial 
to explain, with high accuracy and without any free parameter, the existing
experimental data of the S-factor for the reaction $^{16}$O +
$^{16}$O. More interesting, the radial cranking mass impacts 
significantly on the S-factor excitation function at very low energies, where no
experimental data are available. Clearly, it causes a pronounced maximum 
around 4.5 MeV in the S-factor excitation function. 
To verify this effect new experiments are very desirable. 
In contrast to the results reported in Ref. \cite{Jiang}, the
S-factor does not decline towards extremely low values with
decreasing energy. In fact, below 4 MeV, the S-factor is only
suppressed by a factor of five with respect to the calculation
with the constant reduced mass. The molecular picture discussed in
this work could also be applied to other light systems of great
astrophysical interest that involves deformed nuclei such as
$^{12}$C + $^{12}$C and $^{12}$C + $^{16}$O, provided a general
TCSM for arbitrarily-orientated deformed nuclei is employed. This
work is in progress, and the results will appear in a forthcoming
publication.

% \appendix

%\Acknowledgement{This work was supported by the Joint Institute of
%Nuclear Astrophysics (JINA) through grant NSF PHY 0216783.}

% \label{}
\textbf{Acknowledgement}

This work was supported by the Joint Institute for
Nuclear Astrophysics (JINA) through grant NSF PHY 0216783, and 
by the ARC Discovery grant DP0557065.

\end{document}